\documentclass[11pt]{article}

\usepackage[a4paper,margin=1in]{geometry}
\usepackage{microtype}

\usepackage{amsmath,amsfonts,amssymb}

\usepackage{graphicx}
\usepackage{float}
\usepackage{booktabs,tabularx}
\usepackage{xcolor}
\usepackage{subcaption}
\usepackage{enumitem}
\usepackage{authblk}
\usepackage[hidelinks]{hyperref}
\usepackage{float}  

\title{Gestational Stage Prediction from Cervical Tissue Analysis Using Imaging Mueller Polarimetry Data}

\author[1,${\dagger}$]{Sooyong Chae}
\author[2,${\dagger}$]{Ajmal Ajmal}
\author[2]{Junzhu Pei}
\author[2]{Amanda Sanchez}
\author[2]{Tananant Boonya-ananta}
\author[2]{Andres Rodriguez}
\author[1,2]{Tatiana Novikova}
\author[2,3*]{Jessica C. Ramella-Roman}

\affil[1]{LPICM, CNRS, \'Ecole Polytechnique, IP Paris, 91120 Palaiseau, France}
\affil[2]{Department of Biomedical Engineering, Florida International University, Miami, FL 33174, USA}
\affil[3]{Department of Ophthalmology, Herbert Wertheim College of Medicine, Florida International University, Miami, FL 33199, USA}
\affil[ ]{$^{\dagger}$These authors contributed equally to this work.}
\affil[ ]{*\texttt{jramella@fiu.edu}}
\date{}

\begin{document}

\maketitle

\begin{abstract}
Preterm birth is associated with premature cervical remodeling, yet current clinical assessments cannot detect the underlying microstructural changes in collagen organization. We apply imaging Mueller polarimetry to murine cervical tissue at three gestational stages (Early, mid, late) and develop classification methods to predict gestational stage from polarimetric maps. Using Lu--Chipman decomposition, we extract orientation and azimuth local variability maps that capture collagen fiber alignment and disorder. We evaluate two approaches under 20-fold leave-one-out cross-validation: an analytical threshold classifier on mean azimuth local variability, and a lightweight CNN ensemble (${\sim}76$k parameters) operating on spatially resolved maps. The ensemble achieves 70.0\% sample-level accuracy, outperforming the analytical baseline (55.0\%), with strong performance on early (71\%) and late (86\%) gestation. Spatial prediction maps confirm that classification accuracy is highest in the stroma, where collagen remodeling is most prominent. These results demonstrate that Mueller polarimetry combined with deep learning models can detect gestational collagen remodeling non-invasively, offering a potential pathway toward objective cervical assessment for preterm birth risk.
\end{abstract}

\section{Introduction}
\label{sec:gest_introduction}

Preterm birth (PTB), defined as delivery before 37 weeks of gestation, remains a leading cause of neonatal mortality and morbidity worldwide \cite{koullali2017prevention}. A key indicator of PTB risk is premature cervical remodeling: if the cervix exhibits late-stage microstructural characteristics---such as disorganized collagen---during early gestation, it may signal an elevated risk of preterm delivery. The cervix normally transitions from a rigid, closed structure to a soft, compliant one only near term, driven by progressive reorganization and degradation of collagen fibers in the extracellular matrix \cite{yellon2017contributions, vink2016cervical}. Detecting these microstructural changes early could enable early clinical intervention.

% Current methods for assessing cervical readiness, such as digital palpation and transvaginal ultrasound, are largely subjective or have limited accessibility \cite{feltovich2010quantitative}. They cannot capture the subtle collagen reorganization that precedes mechanical softening. There is therefore a need for non-invasive, quantitative modalities capable of characterizing cervical microstructure at the fiber level.

Imaging Mueller Polarimetry (IMP) has emerged as a promising tool for tissue characterization \cite{wang2021high, he2021polarisation, ramella2022polarized}. By measuring the full $4 \times 4$ Mueller matrix, IMP provides information about depolarization, retardance, and diattenuation---properties directly linked to scattering density, fiber alignment, and anisotropy in cervical tissue \cite{chue2018use, lee2021mueller}. In this study, we use wide-field IMP to investigate collagen remodeling in murine cervical tissue across three gestational stages: early (Day~6), mid (Day~12), and late (Day~18). These time points sample the full remodeling trajectory: Day~6 (D6) represents intact, well-organized collagen; Day~12 (D12) captures the onset of softening; and Day~18 (D18) reflects advanced remodeling immediately before delivery \cite{yellon2017contributions}. We propose both an analytical threshold-based classifier and a CNN ensemble to predict gestational stage from polarimetric maps, evaluated under leave-one-out cross-validation (LOOCV).

\section{Background}
\label{sec:gest_background}

The interaction of a polarized beam with a linear optical system is represented by the matrix equation:
\begin{equation}
\label{eq:lc_MuellerMatrix}
\begin{pmatrix}
S_1^{\text{out}} \\
S_2^{\text{out}} \\
S_3^{\text{out}} \\
S_4^{\text{out}}
\end{pmatrix}
=
\begin{pmatrix}
M_{11} & M_{12} & M_{13} & M_{14} \\
M_{21} & M_{22} & M_{23} & M_{24} \\
M_{31} & M_{32} & M_{33} & M_{34} \\
M_{41} & M_{42} & M_{43} & M_{44}
\end{pmatrix}
\begin{pmatrix}
S_1^{\text{in}} \\
S_2^{\text{in}} \\
S_3^{\text{in}} \\
S_4^{\text{in}}
\end{pmatrix}
\end{equation}

In Eq.~\eqref{eq:lc_MuellerMatrix}, the $4\times4$ real-valued transformation matrix $\mathbf{M}$, known as the Mueller matrix, characterizes how an input Stokes vector $\mathbf{S}^{\text{in}}$ transforms into the output vector $\mathbf{S}^{\text{out}}$ upon interacting with a sample. The Stokes-Mueller formalism is particularly useful for describing the polarization characteristics of partially polarized, fully polarized, or entirely depolarized light, and it is especially effective for modeling interactions with depolarizing medium such as biological tissues.

%--------------------------------------------------------
\subsection{Lu--Chipman Polar Decomposition}
\label{sec:lc_decomposition}
%--------------------------------------------------------

The Lu--Chipman polar decomposition of Mueller matrix is a widely used %framework
non-linear data compression algorithm \cite{lu1996interpretation}.
It decomposes any %experimentally measured
physically realizable Mueller matrix into the product of %physically meaningful polarimetric parameters 
Mueller matrices  %This decomposition models the Mueller matrix as a product 
of three canonical optical elements: a diattenuator $\mathbf{M}_{D}$, a retarder $\mathbf{M}_{R}$, and a depolarizer $\mathbf{M}_{\Delta}$:
\begin{equation}
\label{eq:lcpd-full}
\mathbf{M}
\;=\;
\mathbf{M}_{\Delta}\,\mathbf{M}_{R}\,\mathbf{M}_{D}.
\end{equation}

\subsubsection{Scalar Polarimetric Parameters}
\label{subsec:lc_scalar_polarimetric_paramters}
From the decomposed matrices, scalar polarimetric parameters can be extracted to quantify specific tissue properties. Key parameters include:
\begin{align}
\Delta &\,=\, 1-\tfrac{1}{3}\,\bigl|\text{tr}(\mathbf{M}_{\Delta})-1\bigr|,
\label{eq:depolarization}\\[4pt]
LR &\,=\, \cos^{-1}\!\left(
           \frac{\text{tr}(\mathbf{M}_{R})-1}{2}
           \right),
\label{eq:linear_retardance}\\[4pt]
\psi &\,=\, \tfrac{1}{2}\,
              \tan^{-1}\!\left( \tfrac{M_{R,24}}{M_{R,34}} \right),
\label{eq:azimuth}\\[4pt]
I &\,=\, M_{11}.
\label{eq:intensity}
\end{align}

Here $\Delta$ represents the depolarization (ranging from 0 to 1), $LR$ is the scalar linear retardance and $\psi$ is the azimuth of the optical axis. $I$ is the total intensity (transmitted or reflected).

\subsubsection{Azimuth's Local Variability (ALV)} \label{sec:alv_definition}
To quantify the structural disorder of the collagen network, we introduce the Azimuth's Local Variability (ALV) parameter  \cite{gros2024characterization}. The ALV is derived from the azimuth angle map, $\psi(x,y)$, obtained from the Lu--Chipman decomposition (Eq.~\ref{eq:azimuth}). We compute the circular variance of $\psi$ within a $5\times5$ sliding window centered at each pixel.

For a set of angles $\theta_1, \dots, \theta_N$ within a local window, the mean resultant vector length $R$ is given by:
\begin{equation}
R = \sqrt{\left(\frac{1}{N}\sum_{i=1}^{N} \cos(2\theta_i)\right)^2 + \left(\frac{1}{N}\sum_{i=1}^{N} \sin(2\theta_i)\right)^2}
\end{equation}

%Note the factor of 2, which accounts for the $\pi$-periodicity of the azimuth orientation (i.e., fibers at $0^{\circ}$ and $180^{\circ}$ are equivalent).

The ALV is then defined as the circular standard deviation:
\begin{equation}
\mathrm{ALV} = \frac{180}{2\pi}\sqrt{-2\ln R}
\end{equation}
A value of 0 indicates perfectly ordered alignment (all fibers parallel), while larger values indicate greater local disorder. The scaling factor $\frac{180}{2\pi}$ converts the circular standard deviation to degree-like units.

% The ALV metric ranges from 0 to 1. A value of 0 indicates perfectly ordered alignment (all fibers parallel), while a value of 1 indicates maximum disorder (random orientation).

\subsection{Data Acquisition}
\label{subsec:gest_data_acquisition}

A custom-built wide-field imaging Mueller polarimeter operating at 550\,nm (9W stabilized broadband light source, SLS201L, with FB550-10-1 bandpass filter, Thorlabs) in reflection geometry was used. The system employs polarization state generation (PSG) and analysis (PSA) units to reconstruct the full $4 \times 4$ Mueller matrix of the sample. The PSG of the system consists of a fixed linear polarizer (LPVISC100, Thorlabs) and a quarter-wave plate (AQWP10M-580, Thorlabs) mounted onto a motorized rotational stage (PRM1Z8, Thorlabs). The PSA units contain the same optics as the PSG unit, but in reverse order. The tissues were sampled using a 5x magnifying objective, which was captured by a 16-bit sCMOS camera (PCO.edge 5.5).  

\begin{figure}[htbp]
\centering
\includegraphics[width=0.8\linewidth]{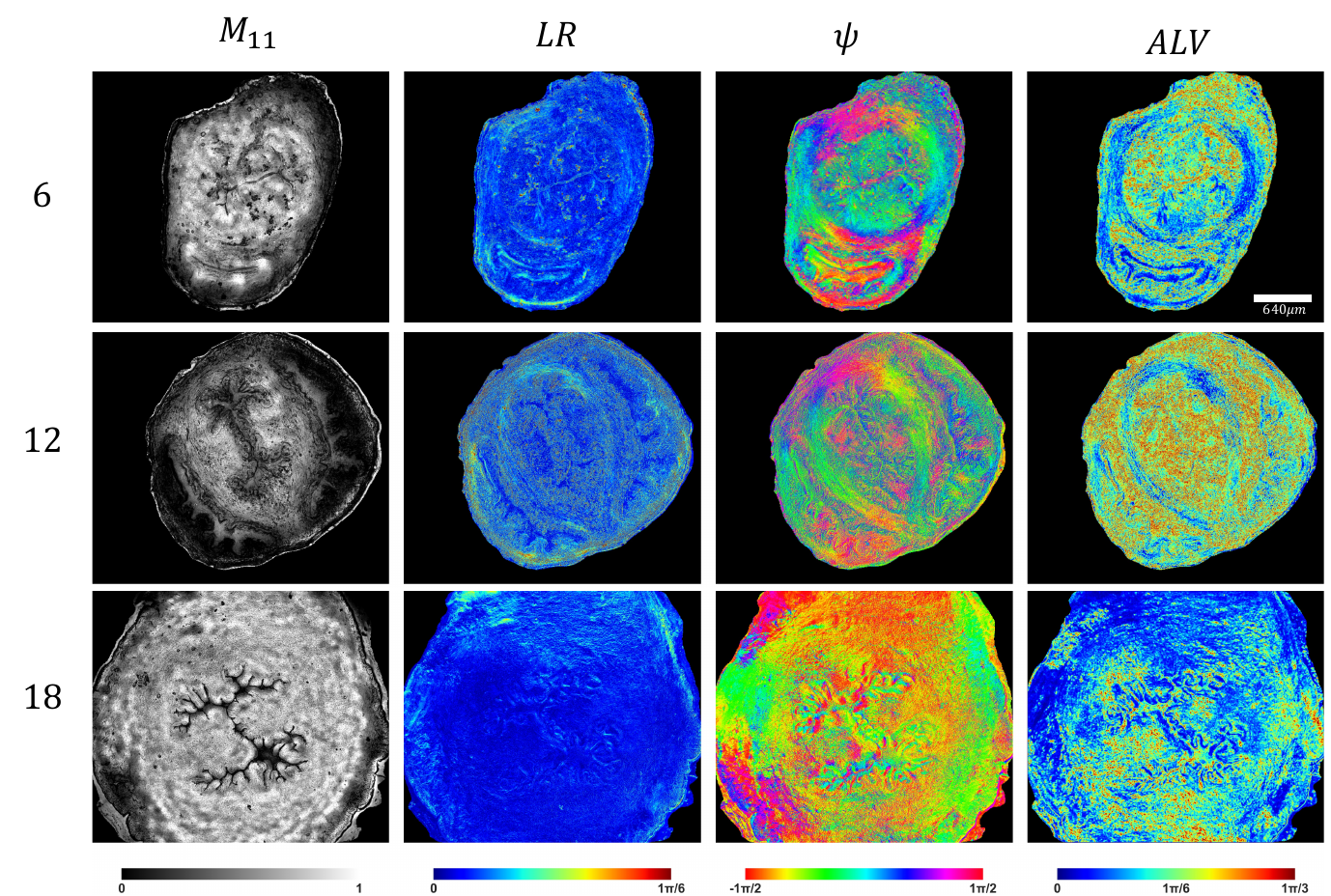}
\caption{Visualization of murine cervix samples across different gestation days: Day 6, Day 12, and Day 18. The tissue regions were filtered using algorithm described in \cite{chae2026intensity}. The widening of the cervical canal is visible in later stages.}
\label{fig:gest_sample_by_day}
\end{figure}

Pregnant female mice, strain C57BL/6J, were used in this study. The start time of gestation (day~0) was determined by observing a copulatory plug. The gestation period for a mouse is typically 19 to 21 days. All animal procedures in this study followed the NIH Guide for the Care and Use of Laboratory Animals and were approved by the Institutional Animal Care and Use Committees at UT Southwestern Medical Center and Florida International University.

Cervices were harvested at D6, D12, and D18, snap-frozen, and cryosectioned at $-20^{\circ}$C into $50\,\mu$m thick transverse sections. Sections were mounted on glass slides without coverslips and imaged within 24 hours of sectioning. The imaging focused on the ectocervix region, which provides a consistent structural landmark across samples. A summary of the dataset characteristics is provided in Appendix~\ref{app:dataset}.

\subsubsection{Region of Interest (ROI) Selection}
\begin{figure}[htbp]
\centering
\includegraphics[width=0.8\linewidth]{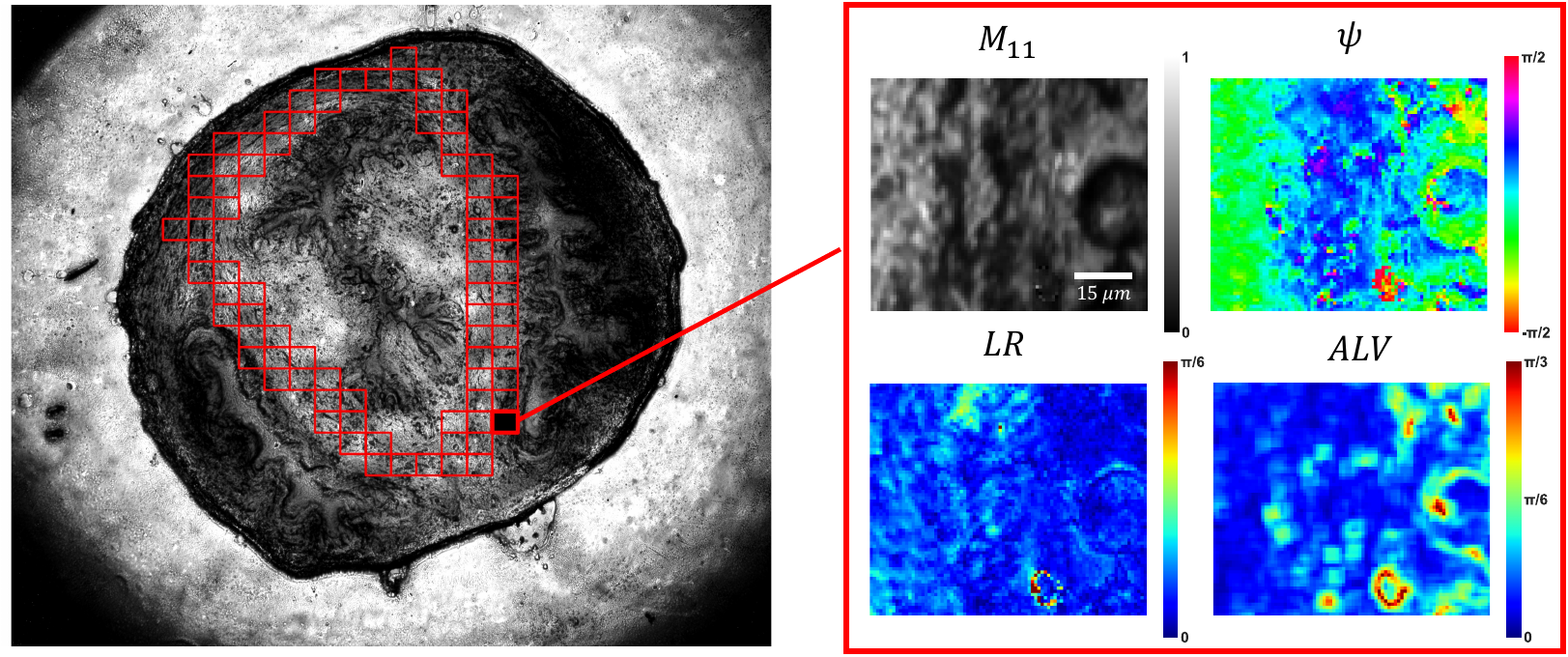}
\caption{Left: M11 intensity image (Day 12) with red contours indicating ROI selection. Right: Expanded polarimetric maps of the selected ROI.}
\label{fig:gest_roi_selection}
\end{figure}

ROIs were selected using a semi-automated grid-based approach. Each specimen image was divided into a $20\times20$ - $35\times35$ grid, and the cells overlying the stromal ectocervix were manually chosen while excluding the surface of the os (OS) and the vaginal wall (Figure~\ref{fig:gest_roi_selection}). Each selected grid cell was extracted as an independent ROI crop at its native resolution and resized to $128\times128$ pixels for model input. A total of 2,844 ROIs were extracted across 20 specimens (see Table~\ref{tab:dataset}).

\section{Methods}
\begin{figure}[ht]
\centering
\includegraphics[width=\linewidth]{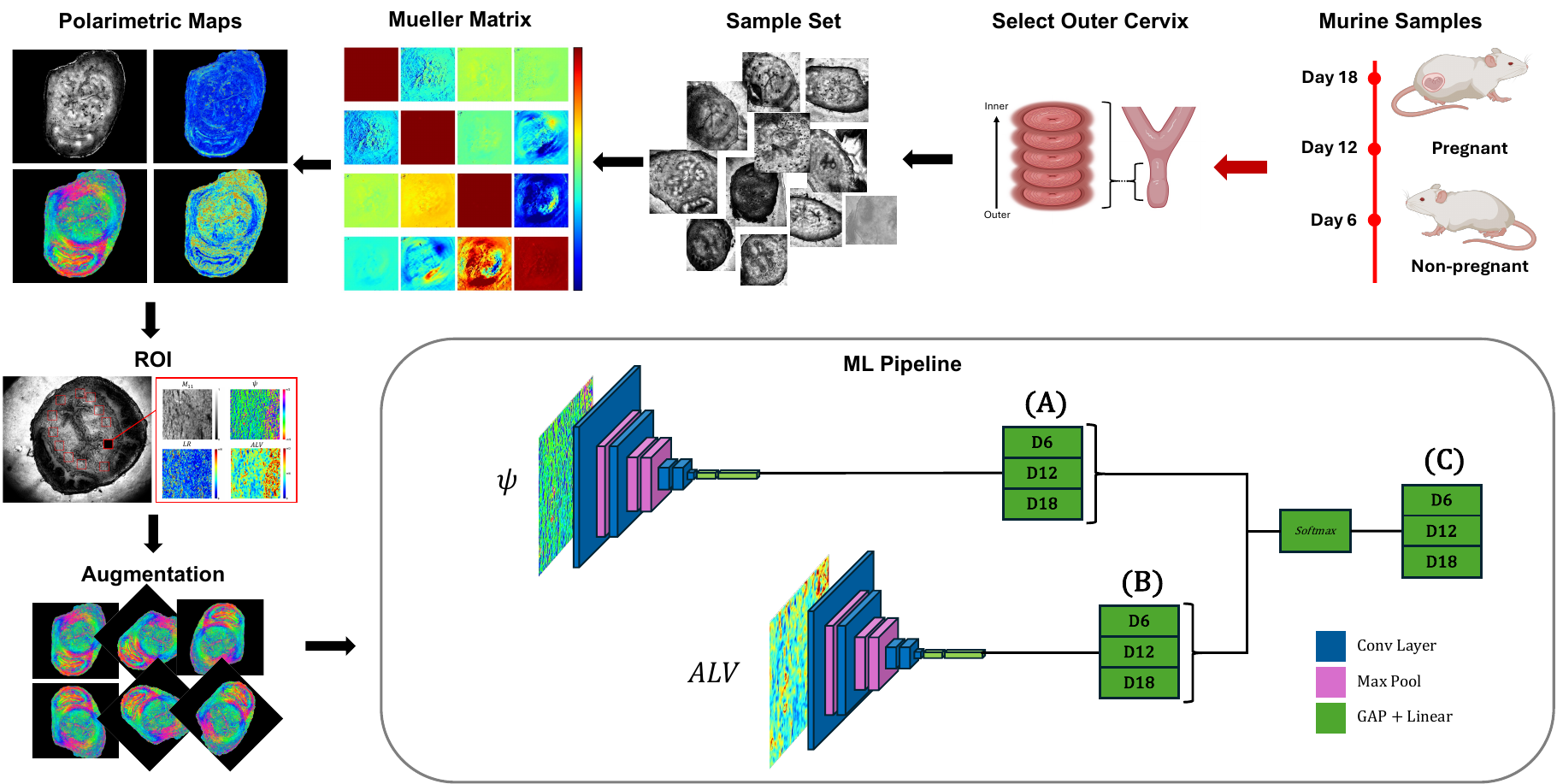}
\caption{Overview of the classification pipeline. Murine cervical tissue samples were collected across gestation, imaged with a Mueller polarimeter, and the resulting polarimetric maps were augmented for training. Two parallel CNN branches operate on (A) the orientation channel and (B) the ALV channel, each with four convolutional blocks (blue) and max pooling (red), followed by global average pooling and a linear classifier (green). Branch outputs are combined by softmax averaging to produce the (C) ensemble prediction.}
\label{fig:gest_workflow}
\end{figure}

We classified specimens into three gestational stages (Day~6, Day~12, Day~18) using two approaches: an analytical threshold classifier based on mean ALV, and a CNN operating on spatially resolved polarimetric maps. Both were evaluated under leave-one-out cross-validation (LOOCV), holding out all ROIs from one specimen per fold so that ROIs from the same section are never split across train and test.

\subsection{Analytical Method: Threshold Classification on Mean Azimuth Variance}

Collagen fiber disorganization increases as the cervix remodels toward delivery \cite{ramella2024quantitative}. To capture this trend with a single scalar, we computed the mean ALV $\bar{ALV}$ for each sample by averaging the pixel-wise ALV (defined in Section~\ref{sec:alv_definition}) over the ROI-selected pixels of each ROI:

\begin{equation}
\label{eq:mean_alv}
\bar{ALV} = \frac{1}{|\mathcal{M}|} \sum_{(x,y)\in\mathcal{M}} V(x,y)
\end{equation}
where $\mathcal{M}$ denotes the set of tissue-masked pixels and $V(x,y)$ is the circular variance at pixel $(x,y)$.

A two-threshold classifier then partitions the samples into three gestational classes based on $\bar{ALV}$:
\begin{equation}
\label{eq:threshold_rule}
\hat{y} =
\begin{cases}
\text{D6}  & \text{if } \bar{ALV} < T_1, \\
\text{D12} & \text{if } T_1 \leq \bar{ALV} < T_2, \\
\text{D18} & \text{if } \bar{ALV} \geq T_2.
\end{cases}
\end{equation}
The decision boundaries $T_1$ and $T_2$ were determined by brute-force grid search over 500 linearly spaced candidates within the observed range of $\bar{ALV}$, selecting the pair that maximized classification accuracy on the training partition. Under LOOCV, thresholds were re-optimized on the training data in each fold to prevent information leakage. This approach was evaluated on the same 20 biological specimens used for the CNN models (7~D6, 6~D12, 7~D18). We report sample-level accuracy from each held-out specimen mean and ROI-level accuracy by applying the fold-specific thresholds to all held-out ROIs.

\begin{figure}[htbp]
\centering

\includegraphics[width=0.65\linewidth]{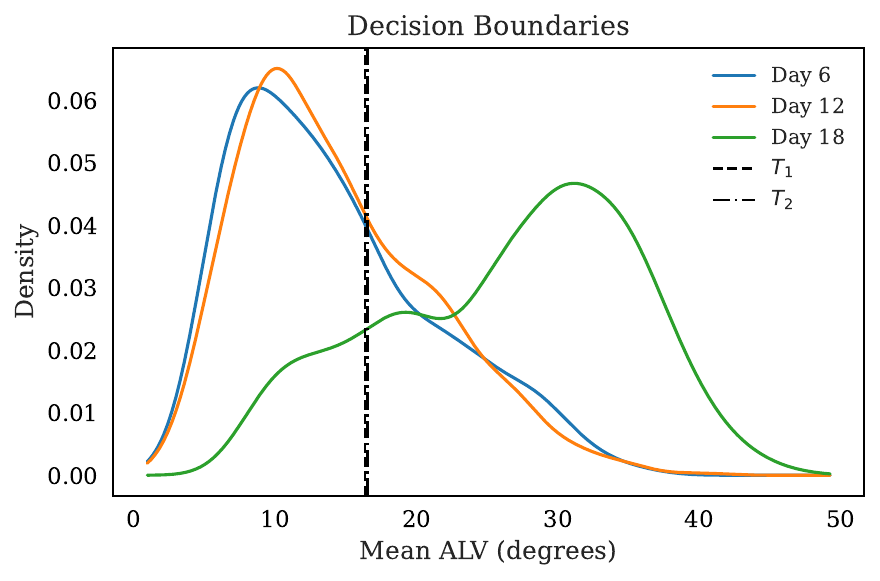}
\caption{Distribution of mean ALV per sample across gestation days, with optimized decision boundaries $T_1$ and $T_2$. D6 and D12 distributions overlap, while D18 is well separated from both groups.}
\label{fig:gest_decision_boundaries}
\end{figure}

\subsection{CNN-Based Classification}
\label{sec:model}

We trained lightweight CNNs on two single-channel inputs derived from the Lu--Chipman decomposition \cite{lu1996interpretation}: the corrected retardance orientation map $\psi(x,y)$ (Model~A), encoding local collagen fiber direction, and the ALV map (Model~B), computed as the circular standard deviation of $\psi$ within a $5\times5$ sliding window. Both models use an identical four-block CNN architecture (LightweightCNN, ${\sim}76$k parameters), described in Appendix~\ref{app:architecture}.

All ROIs were resized to $128\times128$ pixels. For each LOOCV fold, per-channel percentile clipping (1st and 99th) followed by z-score normalization was applied using statistics computed only on the training data. Each training ROI was augmented with one additional randomly transformed copy, resulting in a 2$\times$ expansion of the training set. For Model~A, augmentations comprised random horizontal/vertical flips and $90^{\circ}$ rotations with corresponding angular correction (e.g., $\psi \leftarrow \psi + k\pi/2$, wrapped to $[-\pi/2,\,\pi/2]$). For Model~B, the same spatial transforms were applied without angular correction.

Training configuration and hyperparameters are detailed in Appendix~\ref{app:training}. Both models were evaluated under 20-fold LOOCV; augmented copies were kept in the training partition only. For the ensemble (A+B), the softmax probability vectors from Models~A and~B were averaged per ROI and the predicted class taken by argmax.

\section{Results}
\label{sec:results}

\subsection{Analytical Threshold Classifier}

The mean ALV per ROI increased from D6 through D12 to D18 (Figure~\ref{fig:gest_decision_boundaries}). A Kruskal--Wallis test\cite{kruskal1952use} followed by Dunn's post-hoc \cite{dunn1964multiple} comparisons showed no significant difference between D6 and D12 ($p > 0.05$), while D18 was significantly separated from both groups ($p < 0.01$).

Under 20-fold LOOCV, the analytical threshold classifier achieved 55.0\% sample-level accuracy (11/20) and 57.6\% ROI-level accuracy (1639/2844). At sample level, D18 was classified perfectly (7/7), while D6 (2/7) and D12 (2/6) were frequently misclassified due to the overlap in their variance distributions.

\subsection{CNN-Based Classification}

Model~A (orientation) and Model~B (ALV) each reached 70.0\% sample-level accuracy, with ROI-level accuracies of 62.7\% and 59.6\% (Table~\ref{tab:results}). The ensemble (A+B) reached the same sample-level accuracy and the highest ROI-level accuracy at 65.2\%. The analytical classifier reached 55.0\% sample-level and 57.6\% ROI-level. Per-class results (Table~\ref{tab:per_class}) show 71\% on D6, 50\% on D12, and 86\% on D18 for all CNN models.

\begin{table}[ht]
\centering
\caption{LOOCV classification accuracy for gestation day prediction. All methods were evaluated on the same 20 specimens (2,844 ROIs).}
\label{tab:results}
\begin{tabular}{llcc}
\toprule
Method & Input & Sample (\%) & ROI (\%) \\
\midrule
Orientation (A) & Retardance orientation & 70.0 & 62.7 \\
Az.\ Variance (B) & ALV & 70.0 & 59.6 \\
Ensemble A+B & Prob.\ avg.\ of A \& B & 70.0 & 65.2 \\
Analytical & Mean az.\ variance thresholds & 55.0 & 57.6 \\
\bottomrule
\end{tabular}
\end{table}

\begin{table}[ht]
\centering
\caption{Per-class LOOCV accuracy breakdown (correct/total samples).}
\label{tab:per_class}
\begin{tabular}{lcccc}
\toprule
Class & Orientation (A) & Az.\ Variance (B) & Ensemble A+B & Analytical \\
\midrule
Day 6  & 5/7 (71\%) & 5/7 (71\%) & 5/7 (71\%) & 2/7 (29\%) \\
Day 12 & 3/6 (50\%) & 3/6 (50\%) & 3/6 (50\%) & 2/6 (33\%) \\
Day 18 & 6/7 (86\%) & 6/7 (86\%) & 6/7 (86\%) & 7/7 (100\%) \\
\bottomrule
\end{tabular}
\end{table}

\begin{figure}[!ht]
\centering
\includegraphics[width=0.6\linewidth]{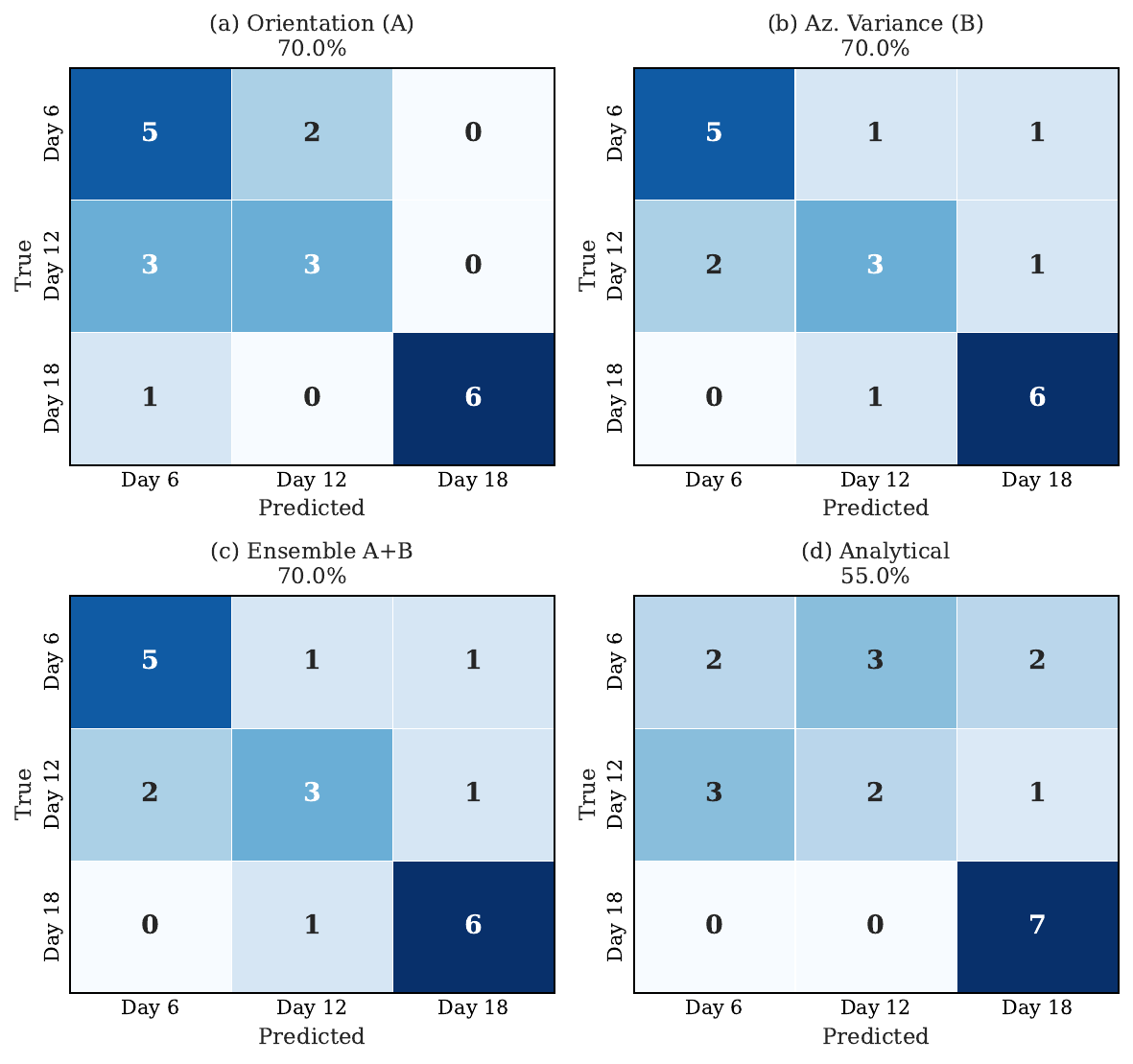}
\caption{Confusion matrices under LOOCV for \textbf{(a)} the orientation model (A),  \textbf{(b)} ALV model (B),  \textbf{(c)} the ensemble (A+B), and  \textbf{(d)} the analytical threshold classifier.}
\label{fig:confusion_matrices}
\end{figure}

Figure~\ref{fig:confusion_matrices} shows the confusion matrices for all methods. In all CNN models, the main errors are between D6 and D12, consistent with their overlapping variance distributions. Model~A predicts the one error on D18 and D6 respectively. The analytical classifier correctly identifies all D18 samples (7/7) but misses D6 and D12.

\subsection{Spatial Prediction Map}

\begin{figure}[H]
\centering
\includegraphics[width=\linewidth]{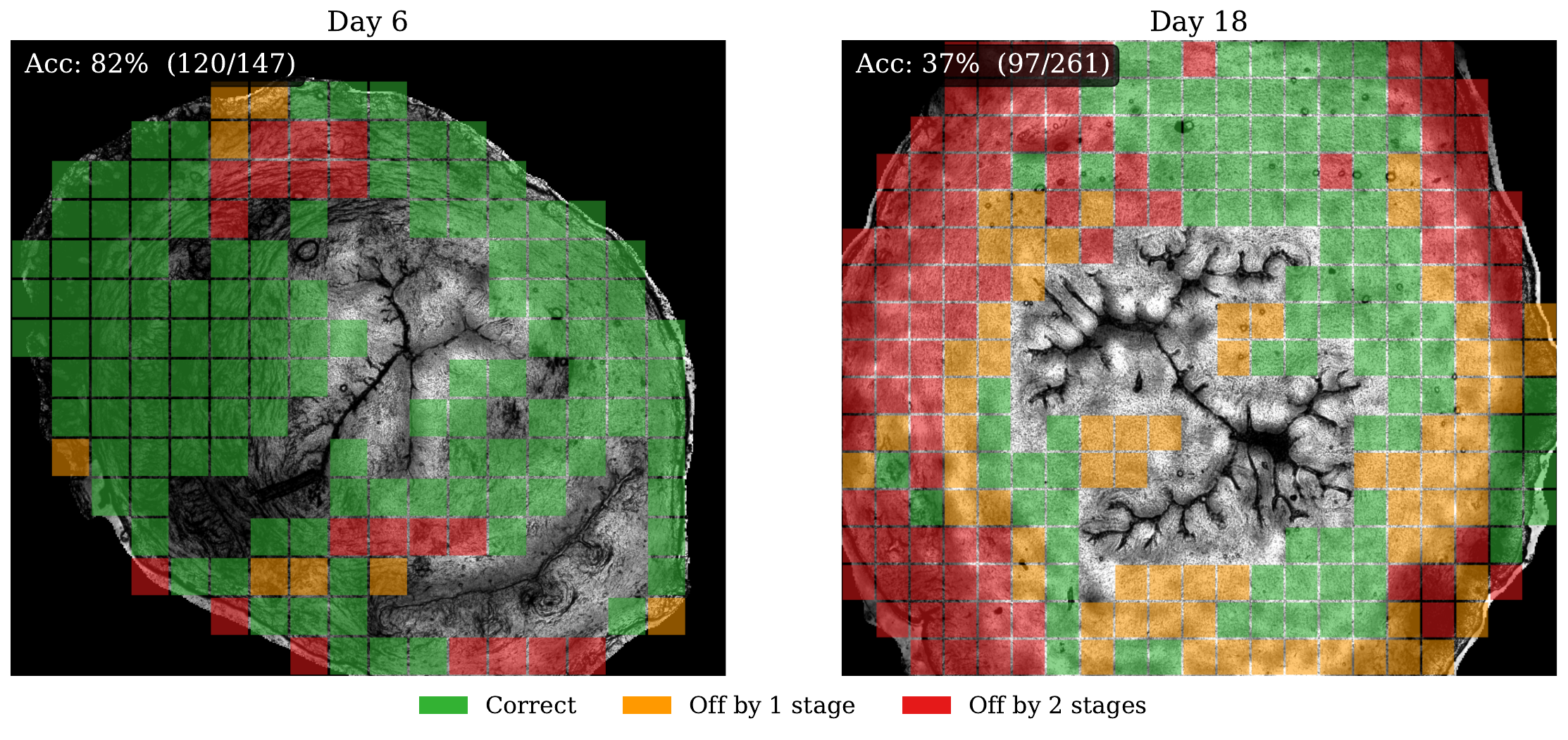}
\caption{Spatial prediction maps for a Day~6 (left) and a Day~18 (right) specimen, masked to the tissue region. Each $60\times60$ pixel patch is color-coded by prediction error: green = correct, orange = off by one stage, red = off by two stages.}
\label{fig:gest_spatial_prediction}
\end{figure}

We constructed spatial prediction maps by tiling the full tissue image into $60\times60$ pixel patches---approximately the size of the ROI grid cells used for training---and running Model~A on each patch independently. Background patches were removed using tissue segmentation \cite{chae2026intensity}, keeping only patches with at least 50\% tissue coverage. Each patch is color-coded by prediction error: green for correct, orange for one stage off, red for two stages off.

Figure~\ref{fig:gest_spatial_prediction} shows the resulting maps for a Day~6 and a Day~18 specimen. For the Day~6 specimen, predictions are mostly correct across the stromal region. In contrast, the Day~18 map is more heterogeneous, with correct and incorrect patches interleaved. Differences in the spatial pattern between specimens are likely driven by sample-level characteristics such as tissue morphology and section quality rather than generalizable model behavior. With only two held-out specimens, there is limited capacity to separate sample-specific effects from true gestational-stage signals.

\section{Discussion}

The CNN ensemble achieved 70.0\% sample-level accuracy under LOOCV, outperforming the analytical threshold classifier (55.0\%). This improvement reflects the CNN's ability to exploit local spatial patterns in fiber organization that are lost when the azimuth variance is collapsed to a single scalar per sample.

Both approaches perform well at distinguishing early (D6) and late (D18) gestation. The ensemble achieves 71\% accuracy on D6 and 86\% on D18 at the sample level, while the analytical classifier reaches 100\% on D18. This confirms that early and late gestational stages exhibit clearly distinct collagen microstructure as captured by Mueller polarimetry, with early-stage tissue showing well-organized fibers and late-stage tissue showing noticeable disorder \cite{ramella2024quantitative}.

A consistent finding across all methods is the difficulty in separating D6 and D12 tissue. The mean ALV distributions of these two groups overlap substantially (Figure~\ref{fig:gest_decision_boundaries}), suggesting that the ectocervix layer does not exhibit a clear change in collagen fiber disorder between early and mid-gestation. This may reflect the fact that ectocervical remodeling at D12 is still in its early stages, with more pronounced structural changes occurring in deeper stromal layers or at later time points. The CNN models partially overcome this by learning spatial texture features beyond global disorder, though D6--D12 confusion remains the primary error mode.

All three CNN models achieve the same 70.0\% sample-level accuracy, with the ensemble reaching the highest ROI-level accuracy (65.2\%) compared to Model~A (62.7\%) and Model~B (59.6\%), suggesting complementary information across the two input channels. The complementary error patterns of the analytical and CNN approaches suggest that a hybrid analytical--ML pipeline could further improve performance. However, it is currently unclear whether the residual errors stem from limitations in ROI selection, the small sample size, or inherent differences between the two approaches. Disentangling these factors would require a larger cohort study.

Mueller polarimetric imaging has potential for translation to bulk tissue imaging through novel hand-held instruments \cite{boonya2024speculum} and has been demonstrated in multiple Mueller polarimetric studies on bulk cervical tissue, although not yet in the context of preterm birth \cite{robinson2023polarimetric, novikova2017optical}.

Despite these limitations, the results establish that Mueller polarimetric imaging combined with lightweight CNNs can detect gestational collagen remodeling in a label-free, non-destructive manner. The approach requires no staining, and the CNN is small enough (${\sim}76$k parameters, ${<}1$\,ms inference per ROI on GPU) for real-time deployment, making it a candidate for future integration into point-of-care cervical assessment tools.

The spatial prediction maps additionally reveal that the OS and vaginal wall regions produce distinct prediction patterns driven by their unique morphological characteristics rather than collagen reorganization. These structures are not part of the current training distribution, yet their polarimetric signatures may carry complementary information about gestational state. Future work could leverage the morphology of these regions, alongside the spatial diversity of stromal predictions, as additional indicators for gestational staging.

\section{Conclusion}

We demonstrated that imaging Mueller polarimetry combined with lightweight CNN classifiers can predict gestational stage from murine cervical tissue with 70.0\% sample-level accuracy under LOOCV. The approach is label-free, non-destructive, and fast ($<$1\,ms inference per ROI on GPU), making it suitable for real-time applications. Spatial prediction maps confirmed that the model performs best in the stroma, where collagen remodeling is the dominant process, validating that the learned features are biologically meaningful. The spatial heterogeneity of predictions observed in late-stage specimens may additionally serve as an indicator of advanced cervical remodeling, a direction worth exploring as larger datasets become available.

The primary limitation is the difficulty in distinguishing early (D6) from mid (D12) gestation, likely because ectocervical remodeling at mid-gestation is subtle at the microstructural level. A larger cohort and imaging of deeper stromal layers may improve this separation. The current dataset of 20 murine samples also limits the precision of accuracy estimates.

From a clinical perspective, the ability to detect late-stage collagen disorganization is the most relevant capability, as it signals advanced remodeling; early detection of such signatures could flag elevated preterm birth risk. Translation to human cervical imaging will require adapting the approach to reflectance geometry and the more complex collagen architecture of human tissue, but the polarimetric features and lightweight model architecture provide a practical foundation for future clinical studies.

\section*{Acknowledgments}
SC, TN acknowledge support from the EUR BERTIP (ANR 18EURE0002, Program France 2030) and European Cooperation in Science and Technology
(COST) actions CA21159 PhoBioS and CA23125 TETRA. JRR acknowledges support from the National Science Foundation (NSF) Award $\#$DMR-1548924. JRR, AA, JZP acknowledge support from the NSF Award $\#$16484510.

\section*{Disclosures}
The authors declare that aspects of the methodology and system described in this manuscript are the subject of a planned patent application. The intellectual property is under preparation and may be filed by Florida International University and collaborators. The authors declare no other conflicts of interest.

\section*{Ethics Approval}
All animal procedures followed the NIH Guide for the Care and Use of Laboratory Animals and were approved by the Institutional Animal Care and Use Committees at UT Southwestern Medical Center and Florida International University.

\section*{Code \& Data Availability}
Code and data are not publicly available at this time due to ongoing intellectual property review, but may be obtained from the authors upon reasonable request.
\appendix

\section{Dataset Characteristics}
\label{app:dataset}

Table~\ref{tab:dataset} summarizes the dataset used in this study. A total of 20 biological specimens were imaged across three gestational stages, yielding 2,844 ROIs selected via the semi-automated grid-based approach described in Section~\ref{subsec:gest_data_acquisition}. Each ROI was resized to $224\times224$ pixels for model input. Each training ROI was augmented with one randomly transformed copy (2$\times$ expansion) per fold.

\begin{table}[ht]
\centering
\caption{Dataset summary. ROI counts refer to original (non-augmented) images.}
\label{tab:dataset}
\begin{tabular}{lccccc}
\toprule
Gestation & Specimens & Total ROIs & ROIs/Specimen & Section & ROI Size \\
Stage &  &  & (range) & Thickness & (resized) \\
\midrule
Day 6 (early)  & 7 & 795   & 55--137  & $50\,\mu$m & $224\times224$ \\
Day 12 (mid)   & 6 & 628   & 82--123  & $50\,\mu$m & $224\times224$ \\
Day 18 (late)  & 7 & 1,421 & 148--248 & $50\,\mu$m & $224\times224$ \\
\midrule
\textbf{Total} & \textbf{20} & \textbf{2,844} & 55--248 & & \\
\bottomrule
\end{tabular}
\end{table}

\section{LightweightCNN Architecture}
\label{app:architecture}

The LightweightCNN consists of four convolutional blocks (each: $3\times3$ Conv2d, BatchNorm, ReLU) with channel progression $1 \to 24 \to 48 \to 64 \to 64$. The first three blocks downsample via $2\times2$ max-pooling; the fourth uses adaptive average pooling to $1\times1$. The resulting 64-dimensional feature vector passes through dropout ($p=0.5$) and a linear layer outputting three class logits. The total parameter count is ${\sim}76$k. This deliberately compact design was chosen because larger architectures (e.g., EfficientNetV2, ResNet) were more prone to overfitting.

\section{Training Configuration}
\label{app:training}

Both models were trained with AdamW (learning rate $10^{-3}$, weight decay $0.01$) using cross-entropy loss with label smoothing ($\epsilon=0.1$) and inverse-frequency class weighting. A cosine annealing schedule reduced the learning rate to $10^{-6}$ over a maximum of 100 epochs, with early stopping (patience~20) based on validation loss. Gradient norms were clipped to 1.0. Within each fold, 15\% of the training samples were held out for validation. Training hyperparameters are summarized in Table~\ref{tab:hyperparameters}. The full 20-fold LOOCV training required approximately 80 minutes per model on a single GPU.

\begin{table}[ht]
\centering
\caption{Training configuration for the CNN-based classifiers
evaluated under leave-one-sample-out cross-validation (LOOCV).}
\label{tab:hyperparameters}
\begin{tabularx}{0.85\linewidth}{lX}
\toprule
Parameter & Value \\
\midrule
Architecture & LightweightCNN (${\sim}76$k params) \\
Batch size & 32 \\
Optimizer & AdamW (lr $= 10^{-3}$, weight decay $= 0.01$) \\
LR scheduler & CosineAnnealingLR ($\eta_{\min} = 10^{-6}$) \\
Epochs (max) & 100 \\
Early stopping patience & 20 \\
Mixup $\alpha$ & 0.2 \\
Label smoothing & 0.1 \\
Validation fraction & 0.15 \\
LOOCV folds & 20 \\
\bottomrule
\end{tabularx}
\end{table}

\bibliographystyle{unsrt}
\bibliography{report}
\end{document}